\documentclass[conference]{IEEEtran}
\IEEEoverridecommandlockouts
\usepackage{cite}
\usepackage{amsmath,amssymb,amsfonts}
\usepackage{algorithmic}
\usepackage{graphicx}
\usepackage{textcomp}
\usepackage{xcolor}

\makeatletter 
\newcommand{\linebreakand}{%
  \end{@IEEEauthorhalign}
  \hfill\mbox{}\par
  \mbox{}\hfill\begin{@IEEEauthorhalign}
}
\makeatother 

\def\BibTeX{{\rm B\kern-.05em{\sc i\kern-.025em b}\kern-.08em
    T\kern-.1667em\lower.7ex\hbox{E}\kern-.125emX}}
\begin{document}

\title{FLoBC: A Decentralized Blockchain-Based Federated Learning Framework}


\author{
\IEEEauthorblockN{Mohamed Ghanem}
\IEEEauthorblockA{
\textit{The American University in Cairo}\\
oscar@aucegypt.edu}
\and
\IEEEauthorblockN{Fadi Dawoud}
\IEEEauthorblockA{
\textit{The American University in Cairo}\\
fadiadel@aucegypt.edu}
\and
\IEEEauthorblockN{Habiba Gamal}
\IEEEauthorblockA{
\textit{The American University in Cairo}\\
habibabassem@aucegypt.edu}
\linebreakand 
\IEEEauthorblockN{Eslam Soliman}
\IEEEauthorblockA{
\textit{The American University in Cairo}\\
eslam98@aucegypt.edu}
\and 
\IEEEauthorblockN{Tamer El-Batt}
\IEEEauthorblockA{
\textit{The American University in Cairo}\\
tamer.elbatt@aucegypt.edu}
\and
\IEEEauthorblockN{Hossam Sharara}
\IEEEauthorblockA{
\textit{The American University in Cairo}\\
hossam.sharara@aucegypt.edu}
}



\maketitle

\begin{abstract}
The rapid expansion of data worldwide invites the need for more distributed solutions in order to apply machine learning on a much wider scale. The resultant distributed learning systems can have various degrees of centralization. In this work, we demonstrate our solution FLoBC for building a generic decentralized federated learning system using the blockchain technology, accommodating any machine learning model that is compatible with gradient descent optimization. We present our system design comprising the two decentralized actors: \emph{trainer} and \emph{validator}, alongside our methodology for ensuring reliable and efficient operation of said system. Finally, we utilize FLoBC as an experimental sandbox to compare and contrast the effects of trainer-to-validator ratio, reward-penalty policy, and model synchronization schemes on the overall system performance, ultimately showing by example that a decentralized federated learning system is indeed a feasible alternative to more centralized architectures.
\end{abstract}

\section*{Graphical Abstract}
\begin{figure}[h]
\centering
\includegraphics[width=.3\textwidth]{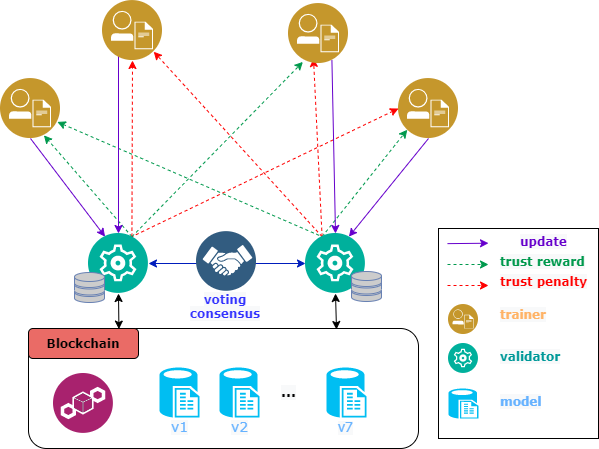}
\end{figure}

\begin{IEEEkeywords}
Byzantine Fault-Tolerance,
Federated Learning,
Blockchain,
Decentralized Systems, 
Distributed Machine Learning,
Privacy Preserving 
\end{IEEEkeywords}
\footnotetext{This paper is the result of BSc in Computer Engineering thesis at the American University in Cairo (Spring 2021)}

\section{Introduction}

\par In recent years, there has been an enormous
corpus of research and development dedicated to
and focused on accelerating the processes of
machine learning in every conceivable shape or
form. Long gone are the days when computer
hardware was not adequate enough to handle
already existing machine learning algorithms, yet
it seems the rapidly growing demand for machine
learning applications has put us in a situation that
yet again undermines the capabilities of individual
machines. Here, an intuitive question comes to
mind: why not just have multiple machines
collaborate on training the model at hand? That is,
in fact, the idea that Google realized in their
conception of Federated Learning (FL) \cite{mcmahan2017communication}. The idea of federated learning solves another major problem: the scarcity of data at a single machine. This is solved through the collaboration of different nodes in the training, each using its local data then sharing its model updates, realizing a single model trained on the sum of their local data. 

In this paper, we present an experimental framework FLoBC for building federated learning systems over blockchain. We further illustrate how it offers feasible solutions to common challenges in decentralized federated learning, namely, \textbf{privacy}, \textbf{efficiency}, and \textbf{Byzantine fault-tolerance}. FLoBC is a proof-of-concept used to spawn mini-systems that demonstrate several aspects of our solution on a relatively small scale, on which we, in later sections, conduct several experiments in order to verify the extent of its efficacy by comparing its performance to a control group for which the features in question are disabled.

\section{Background and Related Work}
\label{similar_work}
 
Prior to the conception of \emph{federated learning}, \textit{blockchain} has been a prominent technology that aimed at realizing decentralized consensus. It is not hard to see the potential merit of combining blockchain with federated learning in order to achieve much more decentralization and privacy. There is a number of contributions in the field of
decentralized federated learning, some of which involve blockchain while others employ other decentralized protocols. The core objective is to eliminate the need for a central server to gather user data and perform model training. The motivation for that is two-fold: on one hand, it provides more privacy, and on the other hand, it relaxes the minimum required amount of computational power by distributing computation across the network. Some of the existing contributions rely on an All-Reduce scheme in which each training node shares its updates with all the nodes in the system resulting in a communication cost of $O(n^2)$ for n training nodes. One important  aspect to leverage (wherever applicable) is network topology which can be
exploited to reduce the communication cost. There are proposals that use the ring topology in \cite{gibiansky2017bringing}, the tree
topology in \cite{li2015malt} and graph topology in \cite{agarwal2014reliable,lalitha2019peer} to share
the model updates of the node with its one-hop
neighbors. However, all these approaches require
multi-hops for the updates to reach all the nodes in
the system, thus leading to slow convergence.

Following the same objective of reducing communication overhead, other contributions, such as \cite{hegedHus2019gossip}, use the gossip protocol which relies on the fact that each peer
sends its updates to another peer in the network, thus propagating the information to the whole network. This arguably puts too much responsibility on the trainer side to validate updates by other peers. To remedy this, our framework introduces validator nodes (similar to cryptocurrency miners) in order to alleviate the computational load of trainers and to have a clearer and easier-to-trace trust system.

Last but not least, and most relevant to our work, some existing work uses blockchain as the communication infrastructure of decentralized federated learning. Blockchain is used to
facilitate uploading and tracking updates, to reward
the users for participating in the training or the
validation of the model, and to make the updates
immutable and secure. In \cite{li2020blockchain}, the role of the
central server in centralized federated learning is
fulfilled by smart contracts. The node performs
the local training, then the local model updates
are sent to an elected consensus committee that
verifies the updates and assigns scores to them. Subsequently, the updated
global model is added to the blockchain. In \cite{kim2019blockchained} and \cite{pokhrel2020federated}, each training node is assigned a validator that validates said trainer's updates, computes proof of work (PoW) and rewards the trainer accordingly. After the updates are added to the blockchain, the miner is rewarded. \cite{martinez2019record} follows a similar algorithm but the trainer can send its updates to any validator and the rewards are given by the model owner, not the validator or the blockchain. \cite{fadaeddini2020secure} and \cite{pokhrel2020decentralized} make use of a distributed peer-to-peer file sharing system to hold the model checkpoints. \cite{pokhrel2020federated, fadaeddini2020secure, pokhrel2020decentralized} are specially tailored for automotive in-vehicle training. \\
By contrast, our framework uses \textit{proof of stake} (PoS) as a much more lightweight alternative to cryptographic PoW. In addition, our methodology is a generic one, hence, it is not bound to a specific task or purpose as long as the required consensus assumptions are met by the underlying network (refer to section \ref{sec:deg_of_cent}).

\section{Methodology}
\label{methodology}

In this section, we introduce our approach for designing a system that closely realizes a decentralized version of \emph{Federated Learning}. In pursuit of that, a few principal standards were set forth to serve as a 
basis for features and solutions employed in the system. These standards are mainly: 
\begin{itemize}
    \item \emph{\textbf{Generality}}: The system design needs to be agnostic of the model specifics apart from it being compatible with gradient descent.
    \item \emph{\textbf{Decentralization}}: System politics shall be undertaken through decentralized consensus.
    \item \emph{\textbf{No Data Sharing}}: Nodes shall only communicate insights, never having to share their data.
\end{itemize}
We cover challenges as well as
solutions pertaining to the performance of
said distributed learning system in the context of
supervised learning where we have a loss function
that we aim to minimize, most prominently
through gradient descent. Though it is mainly
focused on computational aspects, the system also takes into
account communicational efficiency. We
shall discuss the three main dimensions of a
distributed learning system: Parallelism,
Degree of Centralization, and Synchronization.

\subsection{Parallelism}
Training an ML model is, in the great majority of
cases, the heaviest part of the
whole machine learning process \cite{andrychowicz2016learning}, so how could
distributed learning help break this down? For
that, distributed learning generally offers two
paradigms to parallelize the process:
Model-Parallelism and Data-Parallelism \cite{verbraeken2019survey}, the latter of which was chosen for our system.

\textbf{Model Parallelism:}
In model-parallelism, the model structure itself is
split or distributed across multiple nodes, a
paradigm often referred to as Split Learning \cite{kairouz2019advances}.
For instance, in the case of a neural network, the
network would be split at certain layers (called cut
layers), then each trainer would be assigned a group
of consecutive layers, and all layer groups are stitched together based on the layer order in the model architecture \cite{gupta2018distributed}.
While Split Learning has the advantages of data
privacy and enabling nodes with low computation
power to participate in the training, it has some
inherent sequentiality that ultimately limits its
scalability \cite{thapa2020splitfed}. For our purposes, the primary downside of
model-parallelism is that it is model and algorithm
specific, which makes it difficult to utilize for a
general-purpose system for training.

\textbf{Data Parallelism}:
In data-parallelism, each node trains the model in
its entirety, but only on a subset of the training
data, with each node doing this in parallel, then all
generated model updates are aggregated into the global model. This approach has the merit of better
parallel processing on stochastic processes such as stochastic gradient descent (SGD) which is widely used in distributed learning.
Moreover, it can be readily generalized to any
process that lends itself to parallelism such as
gradient computation and approximation \cite{bottou2010large}. As such, we opted for this model of parallelism because it greatly simplifies the general model structure to a flat array of weights, from which the full model can be reconstructed.

\subsection{Degree of Centralization}
\label{sec:deg_of_cent}
It is not hard to see that the two ends of this
spectrum are: fully centralized to fully
decentralized. Accordingly, a strictly fully
centralized system would simply be a single device
doing all of the training process from start to
finish. A more distributed and less centralized
version of that would be the classical
federated learning with a central server
or cluster of servers supervising the training done by other nodes. Here, we present a system on a much higher degree of decentralization. Similar to most cryptocurrency networks, our system comprises two types of nodes: \textbf{trainer} and \textbf{validator}, the latter of which is analogous to a crypto-miner. Validators are responsible for maintaining models on the underlying blockchain network through decentralized consensus based on voting. Ideally, a validator's vote, on whether some update to the model (made by a \emph{trainer} node) is acceptable, should be decided by validation on data presumably unrevealed to the trainer. That, combined with the voting-based nature of consensus, stipulates the following requirements and assumptions in order for the system to operate soundly: 
\begin{itemize}
    \item Strictly more than 2/3 of validators are non-Byzantine.
    \item Different validators' datasets should be sufficiently positively correlated in order to reach a meaningful consensus.
    \item The underlying network is at least partially synchronous.
    \item The quality of models produced by the system is limited by the amount of collective useful data used during training and validation.
\end{itemize}

One
serious implication of the system’s centralization
degree is the choice of the gradient descent (or any
optimization technique) variant that suits how the
system is structured. Since the scope of this study
is focused on distributed learning, we are only
going to consider the distributed centralized
version and its fully decentralized nemesis. As
such, we won’t be concerned with traditional
gradient descent (as it is impossible to use for such
a distributed system), but instead, a stochastic
variant of it should be employed to approximate
the true gradient bearing in mind the rate at which
this approximation converges to the true value
\cite{bottou2010large}. Much like most federated learning systems, our system uses \textbf{\emph{Federated Averaging}} to perform parallel stochastic gradient descent (P-SGD). At its core, it follows a very simple assumption: different nodes training on different datasets
whose updates get aggregated or averaged onto a
single model would eventually converge to a valid
global model.

\subsection{Synchronization}
One crucial aspect of the distributed learning
process is that distributed model updates happen in
tandem to prevent model degradation due to out of
sync parameters \cite{verbraeken2019survey}. Like many other parallel and
massively parallel paradigms, the concept of a
synchronization barrier applies to the case of a
distributed learning system. In this context, each
barrier would signal an update-sharing round in
which updates are aggregated into an updated
model. The more frequently synchronization
barriers occur, the faster the model converges, and
the less it degrades \cite{kairouz2019advances}. However, here lies an
important tradeoff between rate of convergence
and communicational cost. That is, each barrier
incurs a price on the network and node bandwidth.
The main scheme here is that the system has two
types of phases: a computation phase, and
communication phase for sharing the
computational results. For that purpose, there are
many techniques to handle node update
synchronization, and the following subsections
introduce the most common ones.

\textbf{Bulk Synchronous Parallel (BSP):}
Abbreviated as BSP, it is known to be the simplest
approach to preserve consistency by alternating
between rounds of computation followed by
communication. The main pro of this model is that
it has the best consistency (hence, fastest
convergence), yet its main bottleneck is that
finished nodes have to wait for all worker nodes to
finish their computation at each synchronization
barrier \cite{verbraeken2019survey}. In our variant of BSP, each training round is limited by a fixed period that ends with a strict deadline after which no further trainer updates are considered for that round.

\textbf{Stale Synchronous Parallel (SSP):}
This model is similar to BSP but with more relaxed
synchronization barriers that allow a certain
number of iterations beyond the preset number.
After that leeway is reached, all workers are
paused \cite{kairouz2019advances}. Given its nature as a compromise to
BSP, it has good convergence insurances in low to
moderate staleness, beyond which convergence
rates start to decay \cite{verbraeken2019survey}. In our design, SSP is modelled as BSP with the possibility of a deadline extension which is a fraction of the original round period determined in proportion to the ratio of trainers that have not yet finished training for that round, denoted by \emph{slack ratio}. Meanwhile, trainers that already finished by the deadline are allowed to continue training for at most N more steps before the deadline extension finishes.

\textbf{Barrierless Asynchronous Parallel (BAP):}
This model represents the opposite end of the
spectrum to BSP as it almost totally eliminates the
cost of synchronization by allowing waitless,
asynchronous communication between nodes,
which achieves a very low overhead (hence, higher
speedups), but it suffers from potentially slow and
even incorrect convergence with increased delays
\cite{verbraeken2019survey}. In our variant of BAP, trainers are allowed to train for as long as a round may extend; that is, until a new model is released, trainers can keep advancing local training, periodically sharing it with a validator. A new model is released when a minimum ratio of labour has been submitted, and that's when trainers should pull in the new model.

\textbf{Approximate Synchronous Parallel (ASP):}
Although our system does not support ASP, the scheme remains note-worthy. Unlike the SSP model, ASP is concerned with limiting parameter accuracy instead of staleness. It does so by ignoring updates that are not significant to spare the synchronization costs of them \cite{verbraeken2019survey}. However, one downside of that approach is that it is not easy to decide which parameters are insignificant, along with increased complexity of implementation.

\subsection{Byzantine Fault-Tolerance}
Possible cases and scenarios of Byzantine behavior are countless, and decentralized systems are arguably most prone to it. In this work, we focus on lazy\footnote{Lazy, in this context, includes nodes that are not contributing useful updates to models.} and malicious trainer behavior given that malicious validator behavior is mostly taken care of by the blockchain's decentralized consensus. To remedy this, we employ a reward-penalty policy, under which each trainer $i$ is assigned a trust score (or reputation) $\phi_i$ such that:
\begin{center}
    $0 \leq \phi_i \leq 1$
\end{center}
\begin{center}
    $ \sum_{i=1}^{n} \phi_i = 1$
\end{center}
A trainer's trust score is used as the weight factor for that trainer updates in the federated learning algorithm. The effect of that is two-fold. First, it enables validators to control the impact of said trainer's updates on the model based on its trust level. Secondly, it creates intrinsic competition in the system as any rise in one trainer's score effectively results in a relative decline in other trainers' scores. Trust scores are adjusted based on validation results. Upon receiving a trainer update, a validator adds its gradients to the latest model weights, and computes its validation score. Subsequently, it updates the respective trainer's trust score based on whether it leads to improvement or decline.

\section{Architecture}
\label{architecture}

The purpose of this section is to present an overall description of a typical FLoBC\footnote{The word FLoBC is sometimes used to refer to a system spawned through the framework rather than the framework itself.} system such as ones we built to conduct our experiments. Said system was not designed with full deployment in mind, hence some details have been simplified. As such, FLoBC serves as a minimal proof-of-concept.

\subsection{System Overview}
FLoBC is a distributed system driven by two main actors: trainers and validators, the latter of which has more involved responsibilities. On a lower level, the system is composed of six main services, each providing a very specific functionality and has a strictly defined interface with other services and layers.
\begin{enumerate}
    \item \textbf{Blockchain service}: Constitutes the main fabric of communication and data storage across all validator nodes. To ensure data consistency across all validators, this subsystem utilizes an elaborate schema for decentralized consensus, namely a variant of the Practical Byzantine Fault Tolerance (pBFT) family of consensus algorithms that are mainly based on voting and elections.
    \item \textbf{Storage schema service}: Facilitates and enables structured access and modification of the local storage database on which the blockchain state is stored.
    \item \textbf{Machine learning service}: Represents the machine learning layer on top of the blockchain. It is built as a service that handles the execution of machine learning transactions such as gradients sharing by interfacing with the local storage schema to reflect the transaction execution such as model creation and model aggregation.
    \item \textbf{Training validation service}: Is a hybrid layer between the blockchain and the training layer. It is responsible for validating incoming updates and returning a verdict of whether a gradient transaction should be accepted or rejected.
    \item \textbf{Reputation management service}: Handles the accounting aspects of the system, realizing a reward-punishment mechanism to ensure that only honest trainers get to stay in the network while Byzantine ones are dismissed to maintain the performance and quality of created models and ensure fairness.
    \item \textbf{Model training service}: Performs model training including all needed intermediate transformations such as model flattening and rebuilding at the trainer side. This layer sends and receives flattened models and flattened gradients.
    \item \textbf{Training flow management service}: Manages the communication between trainers and validators including model and gradients exchanging, along with synchronizing training iterations.
\end{enumerate}

\subsection{Implementation Technology}
FLoBC is built on top of an Exonum Blockchain that uses a variant of practical Byzantine Fault Tolerance (pBFT) consensus algorithm. The primary reason behind this choice is the voting-based consensus of Exonum being much more lightweight than the typical cryptographic Proof-of-Work (PoW) consensus \cite{yanovich2018exonum}. This lightweight consensus becomes even more crucial when the already computationally heavy nature of model training is taken into consideration. The majority of system services are implemented using the Rust programming language while lightweight clients are implemented using JavaScript. The training and validation are carried out using easily pluggable Python scripts.

\subsection{System Views}
There are multiple ways to view FLoBC depending on the scope level, the most important of which are presented in the following.
\subsubsection{Actor-level View}
This view is the most high-level depiction of the system in terms of its main acting entities and how they are connected.


In our system, we have two types of actors: trainer and validator. While trainers constitute the main work labor in the system by performing gradient calculations using their data, validators are even heavier in composition as they are the ones undertaking elections to achieve consensus and validation to ensure model quality.

One important property of our architecture is that trainers are not fully connected to the validators while validators are fully interconnected. It is worth noting that while the roles (i.e., trainer or validator) are not fixed or dedicated, a node is not allowed to perform both roles simultaneously for the same subnetwork, yet a node can be a validator on a subnetwork while being a trainer for another. Further, each model subnetwork requires at least one validator to function.

\subsubsection{Modular \& Service-level Views}
\begin{figure}[h]
    \centering
    \includegraphics[width=.46\textwidth]{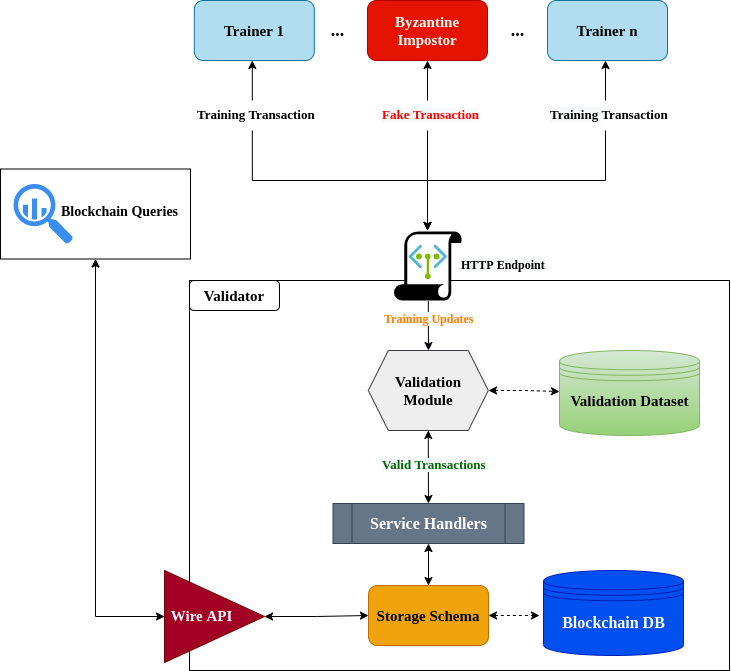}
    \caption{Modular View}
    \label{fig:byzantine_modular_view}
\end{figure}

\begin{figure}[h]
    \centering
    \includegraphics[width=.3\textwidth]{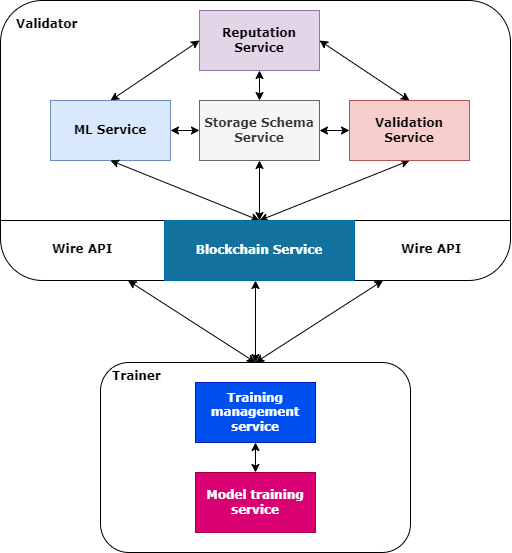}
    \caption{Service-level View}
    \label{fig:service-level}
\end{figure}

The modular view in Figure \ref{fig:byzantine_modular_view} gives a somewhat closer look on a validator's insides, showing the data flow from each high-level module to the others. It first shows that trainers communicate with validators over HTTP whose endpoint then passes the transactions to the validation module that uses the validation dataset to validate the model-update transactions in order to ultimately decide whether to accept of reject the update onto the blockchain. Validators also expose a Wire API for answering queries about the blockchain state (e.g., what's the latest model version?), which is helpful for system diagnosis/monitoring. On the other hand, the service-level view emphasizes the separation of concerns/responsibilities across the system into multiple services illustrated in Figure \ref{fig:service-level}. The storage schema service is central to all services in the validator side as it represents the main interface for making persistent changes. The ML service is responsible for consolidating trainer updates and handling the trainer flow (e.g., synchronization) while the validation service is responsible for assessing trainer work and providing the results to the reputation service which can adjust trust scores accordingly. The Wire APIs on the sides represent auxiliary querying interfaces for miscellaneous purposes (e.g., for getting the latest released model version) besides the main blockchain interfaces used for sending transactions.

\section{Experiments}
\label{experiments}
In this section, we describe the experiments we have carried out using FLoBC. All the experiments carried out in this section were training and validating a convolutional neural network (CNN) predicting handwritten MNIST digits on the regular MNIST dataset with images of size $28\times 28$ pixels. Each trainer in the system uses a random 30\% sample of the MNIST training dataset. The metric used to evaluate training in these experiments is accuracy, but the system will work equivalently using any machine learning model metric. The metric is easily pluggable in the Python machine learning validation script.

\subsection*{\textbf{Benchmark: Decentralized vs. Centralized Performance}}
In this experiment, we compare the performance of the centralized model with that of the decentralized model utilizing the best performing configuration (as demonstrated later by experiment 1) comprising 7 trainers and 3 validators with an active reward-penalty policy. The centralized model is the golden benchmark of our system. The aim of this experiment is to test whether the added benefits of decentralization would result in a significant cost of model quality.
\subsubsection*{\textbf{Setup}}
In this experiment, the centralized model uses the whole data available in the MNIST training data-set. Since in the best performing decentralized configuration there are 7 trainers affecting the model each iteration, in our centralized benchmark, each training iteration is 7 epochs. The training lasts for 30 iterations for both the centralized and decentralized runs.

\subsubsection*{\textbf{Results and Discussion}}

\begin{figure}[h]
    \centering
    \includegraphics[width=.5\textwidth]{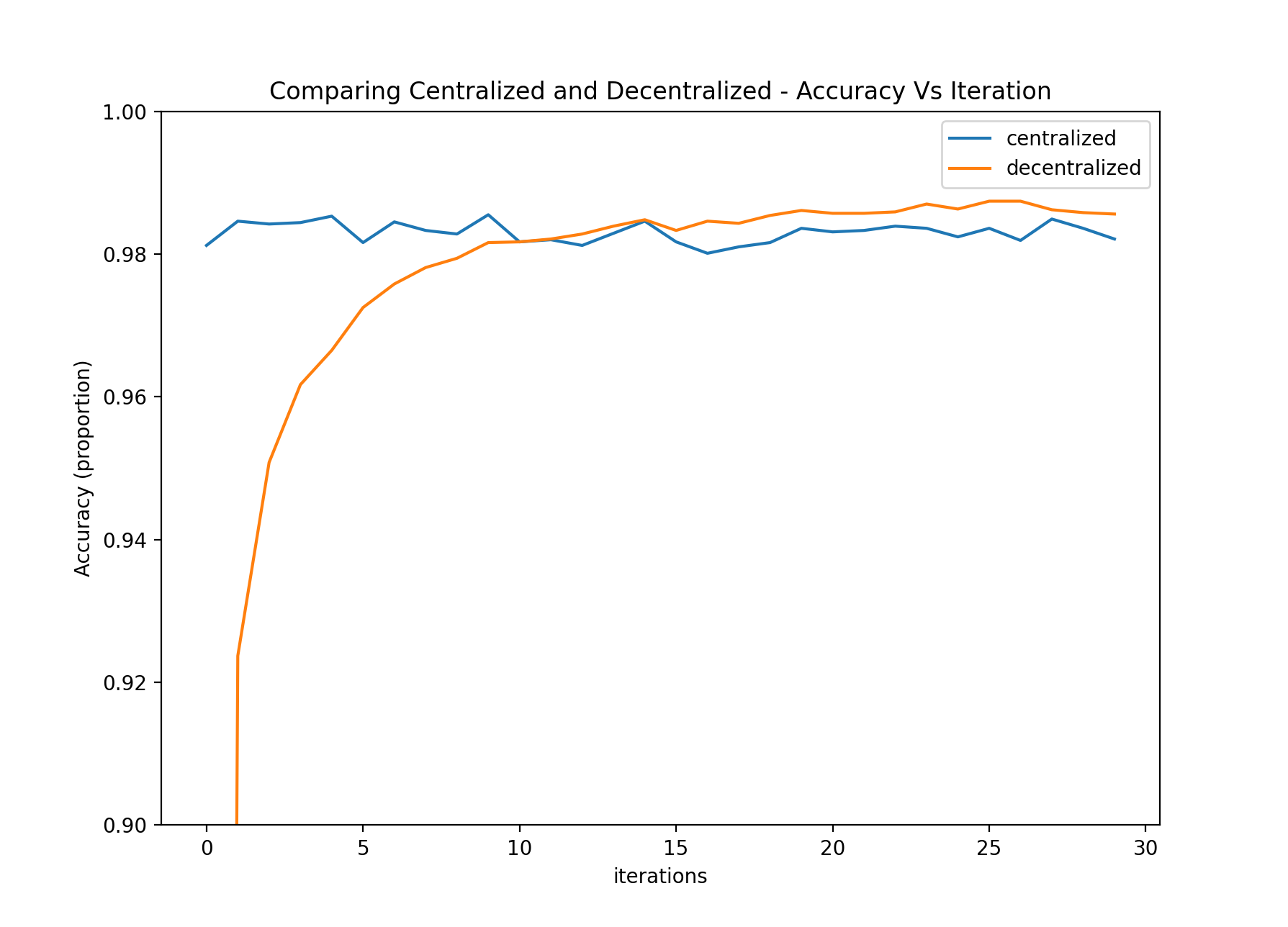}
    \caption{Accuracy against iterations for the centralized and decentralized runs}
    \label{fig:exp4}
\end{figure}
As it can be seen in Figure \ref{fig:exp4}, the centralized model had a higher accuracy than the decentralized alternative for 10 iterations, after which the decentralized model trained by 7 trainers and validated by 3 validators marginally outperformed the centralized model. The difference in accuracy is below 0.5\%. This experiment shows that the added benefits of decentralization and privacy preserving training do not degrade the model performance. Note that this is not meant to demonstrate that the decentralized version is superior in performance to its centralized counterpart. Rather, it shows that the two are closely comparable in that regard. 
\subsection*{\textbf{Experiment 1: Trainers-to-Validators Ratio}} 
In this experiment, we have maximum N participating nodes. Each run we split N into different numbers of trainers and validators. Through this experiment we want to determine the best trainers-to-validators split that makes use of the most computing recources and achieves the best model performance while still relying on multiple validators to generate a more trustworthy model. This experiment will also show the speed of convergence to the highest accuracy by pointing out the iteration at which the highest accuracy was achieved. There is a trade-off between the number of trainers and validators that this experiment aims to balance out. Note that  increasing the number of trainers with good quality data should increase the model performance. Greedily, we would want all the nodes in the system to be trainers, however, this has a major drawback. Relying on a single validator means that the validator is a trusted entity whereas relying on multiple validators increases the trustworthiness of the model because of consensus. However, we want to refrain from having too many validators because this wastes computing resources that could have been utilized in training and adds on unnecessary communication costs. A trainer shares its updates with one validator that validates those updates then shares them with the rest of the validators for further validation and finally consensus. This being said, as we increase the number of the validators, we increase the communication cost associated with voting consensus. For these reasons, for the highest efficiency and the best performance, it is important to determine the best trainers-to-validators ratio that achieves the best model quality without compromising decentralization. 
\subsubsection*{\textbf{Setup}}
In this experiment, the N number of agents in the system is set to 10. For instance, with N = 10, one system run has 9 trainers and 1 validator while another run has 8 trainers and 2 validators, etc. Each run, we change the number of trainers and validators to try all their combinations that add up to 10 agents. Each run lasts for 30 iterations. We record the iteration and its accuracy for each run of the system.
The new model version creation waits for all the trainers to share their updates with the validators since it is important in this experiment that all trainers submit their work for validation each model update iteration since we are studying the effect of the split of the validators and trainers on the model performance.

\subsubsection*{\textbf{Results and Discussion}}

\begin{figure}[h]
    \centering
    \includegraphics[width=.5\textwidth]{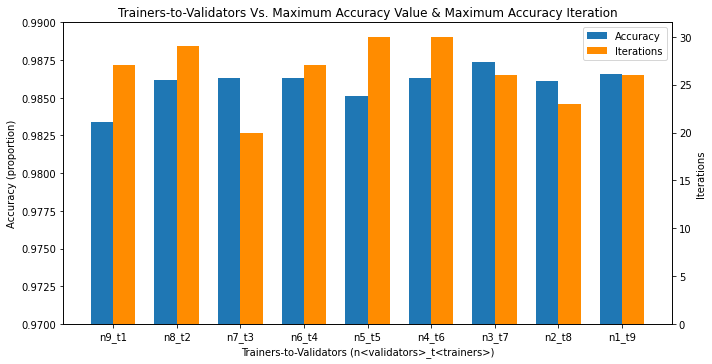}
    \caption{Maximum accuracy and its iteration index for different splits of trainers and validators}
    \label{fig:exp1}
\end{figure}
As it can be seen from Figure \ref{fig:exp1}, the best configuration for this model is having 3 validators and 7 trainers. Using this configuration, the maximum accuracy of 0.9874 was reached at iteration 26. Note that the least accuracy was that for the single trainer configuration since much less computing resources and data have been dedicated for training.  
\subsection*{\textbf{Experiment 2: Reward-Penalty Policy}} 

The purpose of this experiment is to evaluate the usefulness of computing a score for each trainer, where a trainer's score affects the extent to which this trainer's updates affect the overall model.

\subsubsection*{\textbf{Setup}}

This experiment was performed using processes representing 6 trainers (indexed from 0 to 5) and 3 validators. Training ran for 30 iterations. The BSP synchoronization scheme was employed with a sufficiently large period to allow all trainers to submit updates every synchronization period.

The updates of each of the trainers were offset using an approximately normal noise. The mean of the noise is 0 and the standard deviation of the noise is linearly proportional to the index of the trainer, with proportionality constant k = 0.0545. Thus, the updates of trainer 0 received zero noise, and the updates of trainer 5 were significantly offset by the noise.

For purposes of this experiment, the minimum acceptance threshold for updates was reduced substantially to allow low-performing updates to affect the model generation and be affected by scoring. 
The control group had scoring disabled; that is, the scores of all 6 trainers were constant and equal throughout the training process. The scoring group allowed scoring for each trainer to change and to adapt to the relative qualities of the provided updates by each trainer.

\begin{table*}[t]
\caption{Scores of all trainers across training rounds}
\centering
\begin{tabular}{|c|c|c|c|c|c|c|c|}
\hline
Round     & 0       & 5        & 10       & 15      & 20       & 25      & 30       \\ \hline
Trainer 0 & 0.16667 & 0.764622 & 0.823812 & 0.866141 & 0.907603 & 0.964613 & 1 \\ \hline
Trainer 1 & 0.16667 & 0.235378 & 0.176188 & 0.133859 & 0.092397 & 0.035387 & 0 \\ \hline
Trainer 2 & 0.16667 & 0        & 0        & 0       & 0        & 0       & 0        \\ \hline
Trainer 3 & 0.16667 & 0        & 0        & 0       & 0        & 0       & 0        \\ \hline
Trainer 4 & 0.16667 & 0        & 0        & 0       & 0        & 0       & 0        \\ \hline
Trainer 5 & 0.16667 & 0        & 0        & 0       & 0        & 0       & 0        \\ \hline
\end{tabular}
\label{table:scores}
\end{table*}

\subsubsection*{\textbf{Results and Discussion}}

During the operation of the scoring group, the scores of each trainer were updated after every synchronization period. A sample of the scores of all trainers over the 30 training iterations is displayed in Table~\ref{table:scores}. We note that, shortly after training started, the validators realized that most trainers were providing low-quality updates, and thus, the scores of four trainers dropped to zero almost immediately. Over the remaining rounds, the score of trainer 0, whose updates were not affected by added noise, steadily increased against the score of the second-best trainer, trainer 1, whose updates were affected by a minimal amount of noise. Hence, we conclude that scoring enables the validators to accurately estimate the relative performance of each of the trainers.

\begin{figure}[h]
    \centering
    \includegraphics[width=.45\textwidth]{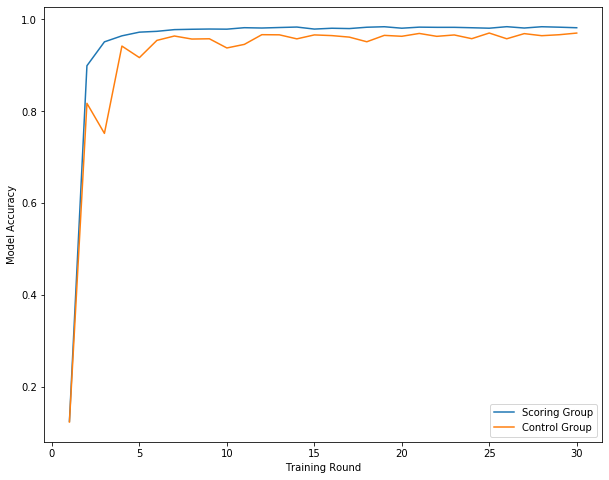}
    \caption{Accuracy performance of scoring group vs control (uniform) group over 30 training rounds}
    \label{fig:exp2}
\end{figure}

For the two groups, after each training round, the current model accuracy is computed. The accuracy performance of the two groups is shown in Figure~\ref{fig:exp2}. We note that trainer scoring improved the performance, stability, and convergence speed of the model, as only the updates from high-performing trainers were taken into consideration and contributed to improving the model.

\subsection*{\textbf{Experiment 3: Synchronization Schemes}} 

In this experiment, we compare some of the common synchronization schemes used in parallel computing after adapting our own variations of them. The schemes implemented in FLoBC and those tested in the experiment are Bulk Synchronous Parallel (BSP), Stale Synchronous Parallel (SSP), and Barrierless Asynchronous Parallel (BAP), as previously described in the methodology. The purpose of the experiment is to compare the different schemes in terms of two main metrics: model growth relative to the number of training rounds, and the average time per training round. The former indicates convergence while the latter indicates progression speed.

\subsubsection*{\textbf{Setup}} 

For the experiment setup, a system of three validators and six trainers is used in all runs with a uniform trust policy; that is, all trainers have the same trust scores. For both BSP and SSP, the same base synchronization barrier period is used. Different trainers are running at different paces to emulate real-world computational differences, with the synchronization barrier period set to be longer than the time most (but not all) trainers will take to finish one training job. To measure the model growth metric, the system will run for a fixed N=30 training rounds, comparing how long each scheme takes to converge on the highest model accuracy while also taking the three highest accuracies into consideration. As for the average training iteration delay, the system will run for a fixed period of 20 minutes. For BAP, we use two configurations: one with a slack ratio threshold of 0\% (fully relaxed), and another with 40\%. That is, in the first one, a sync barrier is thrown when all trainers have submitted at least one update in the current training round while the second one is contingent on only 60\% of trainers having done so.
\vspace{-1.5em}
\begin{figure}[h]
    \centering
    \includegraphics[width=.5\textwidth]{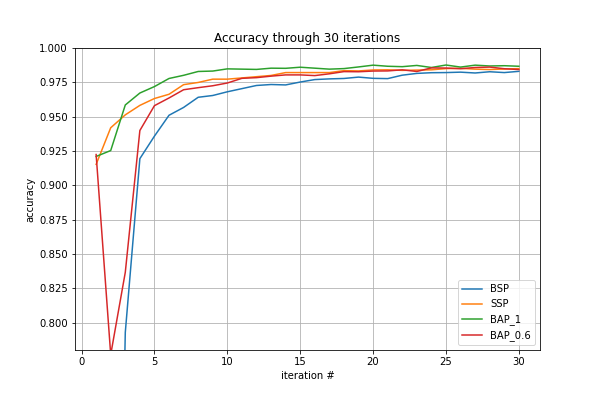}
    \caption{Accuracy performance for 4 different schemes across 30 training iterations}
    \label{fig:exp3_30iterations_4schemes}
\end{figure}

\subsubsection*{\textbf{Results and Discussion}}

As depicted in Figure \ref{fig:exp3_30iterations_4schemes}, results seemingly meet their theoretical expectations, confirming the trade-offs for the three different schemes. BSP has the lowest performance as it is the most strict in terms of synchronization, leading to lower utilization of training power. However, it is quite stable as it offers more predictable limits on speed of progression (at a fixed period). As for the two BAP runs\footnote{Labelled as BAP$\_<$majority ratio$>$}, BAP\_1 with a majority ratio of 100\% was expected to perform best since it has the best utilization of trainer work, beating its BAP\_0.6 counterpart, which confirms the expected effect of decreasing the majority ratio in our variant of BAP. On the other hand, SSP appeared to strike a balance between BAP and BSP. This is expected since SSP is regarded as a compromise between strict synchronization (BSP) and full relaxation (BAP). 

\begin{figure}[h]
    \centering
    \includegraphics[width=.5\textwidth]{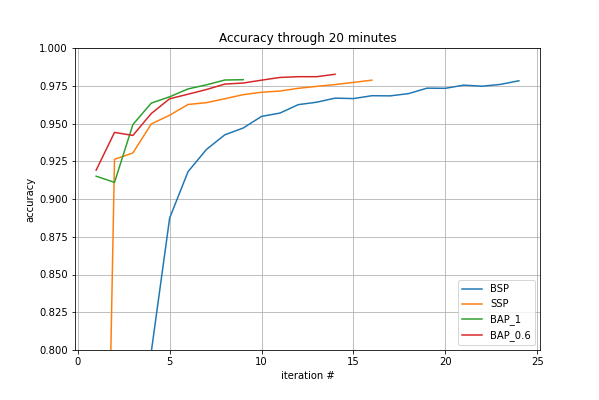}
    \caption{Accuracy performance for 4 different schemes across 20 minutes of training time}
    \label{fig:exp3_20minutes_4schemes}
\end{figure}

Looking at Figure \ref{fig:exp3_20minutes_4schemes}, fully relaxed BAP performed the least number of iterations (i.e., training rounds) by far due to its relaxed sync barriers. Naturally, partially relaxed BAP managed to perform significantly more rounds, eventually leading it to score the highest accuracy of all schemes. On the other hand, BSP had the furthest progression, managing to score very close to the top accuracy due to its faster pace. SSP performed fewer training iterations than BSP since it applies more relaxation on the sync barrier. However, SSP still outperforms BSP in terms of accuracy due to its better utilization of the training power. By and large, each one of these schemes suits certain system configurations (and purposes), depending on several factors, but the primary deciders are trainer pace and quality. 


\section{Conclusion}
\label{conclusion}
In summary, FLoBC is a proof-of-concept that is intended to show that its components can come together to form a coherent system with the desired levels of generality, decentralization, efficiency, privacy, and Byzantine fault-tolerance. Through our benchmark experiments, our decentralized system has shown itself to be a viable match to pure centralized training, marginally outperforming the centralized benchmark by a factor of 0.5\%. Using toy systems spawned by FLoBC, we first showed that there is a quasi-optimal balance to strike on trainer-to-validator partitioning, empirically determining that a 7:3 trainer-to-validator ratio works best. Secondly, we demonstrated that even a simple reward-penalty policy can have a notable positive effect on the quality of produced models. Finally, we compared three major synchronization schemes (BSP, SSP, \& BAP), highlighting and contrasting their different trade-offs.
\section{Future Work}
\label{future_work}
Despite the SGD-compatibility of a model being an assumption embedded in the system, FLoBC can be feasibly extended to handle other computational models. 
Further improvements can be applied to synchronization schemes  by allowing more adaptability in the sync periods based on the expected time that most trainers take to submit their updates in one round. Furthermore, to ensure a higher level of data privacy, a layer of differential privacy could be incorporated into transaction communication to limit the extent to which gradients can be inverted to extract trainer data.


\bibliographystyle{plain}


\bibliography{ref}
\end{document}